\documentstyle[11pt,dunk2001_asp,twoside,epsf]{article}
\def\edcomment#1{\iffalse\marginpar{\raggedright\sl#1\/}\else\relax\fi}
\reversemarginpar

\begin{document}
\title{On The Origin of HI in Galaxies:\\
The Sizes and Masses of HI Photodissociation Regions}
\author{Ronald J. Allen}
\affil{Space Telescope Science Institute, 3700 San Martin Drive,
Baltimore, MD 21218, USA}

\begin{abstract}
Young stars in the disks of galaxies produce HI from their parent
H$_2$ clouds by photodissociation. This paper describes the observational
evidence for and the morphology of such HI\@. Simple estimates of the
amount of dissociated gas lead to the startling conclusion that much,
and perhaps even all, of the HI in galaxy disks can be produced in
this way.
\end{abstract}

\section{Introduction}

Most observers view the clouds of HI we see in galaxies as the raw material
out of which the stars were formed.  These clouds are thought to form
higher-density complexes of gas and dust, turn molecular (H$_2$), and then
form stars.  In this context, the observed correlation between the star
formation rate in galaxy disks and the HI content (often called the Schmidt
Law) is generally viewed as being at the basis of an understanding of the
global star formation process in galaxies.  However, after many years of
work to elucidate the specific physics of this process, we have a
complicated and still rather rudimentary picture of just how all this might
happen.

I want to propose a different view for a part of this ``star-formation''
story.  This view is contrary to the conventional wisdom, but it has the
virtue of being physically much simpler.  I hope to convince you that this
view is supported both qualitatively, by the detailed morphology of the HI
observed in galaxy disks, as well as quantitatively, by the theory of
photodissociation of H$_2$.  In this view, the basic star construction
material in galaxies is gas which is already mostly molecular, and out of
which the stars form directly.  HI appears in the region when the leftover
H$_2$ is illuminated with UV photons from nearby young stars.  The physics
of photodissociation regions provides a natural and quantitative
explanation for the appearance of HI envelopes around the clouds, and for
the CO(1-0) emission which is sometimes seen emanating from the warmer,
higher-density parts of their surfaces.  The rather surprising new result
is that the total amount of Far-UV emission produced in galaxy disks by
run-of-the-mill, non-ionizing B stars is actually sufficient to account for
most, and perhaps even all, of the HI present.

\section{Background}

For me, the first example of failure of the conventional wisdom about HI
being the precursor to star formation appeared in the early VLA-HI
synthesis images of the relatively nearby galaxy M83, which I first saw at
an afternoon seminar given by Mark Ondrechen at the Kapteyn Laboratory in
Groningen in 1985.  The morphology of the HI in the spiral arms of this
galaxy did not fit the expected picture; rather than coinciding with the
dust lanes as markers of the highest gas column density, the HI ridges
appeared to be shifted to larger galactocentric distances and to correspond
better with the line of HII regions marking the locus of the youngest stars
in the spiral arm.  The density wave picture for spiral structure
introduced by Lin \& Shu (1964) led to a simple explanation (Roberts 1969)
for the separation of the arm of maximum gas column density (located
upstream in the flow) and the HII regions (downstream) in terms of the flow
velocities and the time to convert the H$_2$ in GMCs into stars.  But if
the HI was a precursor to the H$_2$ in GMCs, this picture did not explain
why the HI ridge also appeared downstream with the HII regions.  The
explanation we offered (Allen, Atherton, \& Tilanus 1986) was that the ISM
was changing its physical state as it moved along, and doing so in a big
way.  Primarily molecular to start with, the residual gas after star
formation becomes atomic as it is bathed in UV photons from the most
massive of the nearby newly-formed stars. It later returns to the molecular
form after the young stars die off and the UV flux decreases.

Other studies have followed on M83 and on other nearby spirals (M51, M100)
and have generally reached the same conclusions; a list of relevant
references is given in the introduction to the paper on photodissociation
in M101 by Smith et al.\ (2000); that introduction also summarizes the basic
physics and provides many additional references to the literature on the
physics of Photo-Dissociation Regions (PDRs).

With the exception of the most recent work on M101, the studies referenced
above have been carried out on relatively coarse linear scales, typically
500 - 1000 pc.  This is too coarse to even begin to resolve the morphology
of individual PDRs, which form ``blankets'' and ``blisters'' on the
surfaces of GMCs with the largest scales of $\sim 100$ pc.  Such structures
were first identified in M81 with a resolution of $\sim 150$ pc (Allen et
al.\ 1997) and in M101 with a resolution of $\sim 220$ pc (Smith et al.\
2000).

In this paper I want to start from the ``other'' end, at length scales of
$\sim 1$ pc, and explore the nature of the HI structures associated with
GMCs in the Galaxy.  These structures can also be explained with the
photodissociation model, and provide new information on how to quantify the
relation between UV-producing B stars and HI\@.  To give the story some
structure I will describe the examples on scales of a factor of 10 from 1
to 1000 pc, providing overlap with the studies described above and showing
the continuity and ubiquity of the photodissociation process in the ISM of
galaxies.

\section{HI associated with H$_2$ on the $\approx 1$ pc scale}

On the scale of $\approx 1$ pc the radio -- HI resolution is too poor to
look for HI -- H$_2$ associations anywhere except in the Galaxy, and there,
confusion along the line of sight usually makes the identification of HI
features with specific stars and GMCs very difficult.  Nevertheless, new
all-sky HI surveys at high resolution have revealed several cases of
discrete HI features associated with young stars deeply embedded in dense
molecular clouds.

\subsection{Deeply-embedded young stars}

The prototype of this class is IRAS 23545+6508 (Dewdney et al.\ 1991).
This object is a strong IR source and a weak radio continuum source, and
also has a faint reflection nebula apparently associated with it.  The HI
source is compact, $\approx 0.6$ pc in diameter, with M(HI) $\approx 1.4$
M$_{\odot}$.  The star is B3 or B4 but behind $A_V \approx 11$ mag of
extinction, embedded in a dense GMC at $\approx 1$ kpc distance.

Several other examples like IRAS 23545+6508 have been found (Purton,
private communication), including an object associated with the star BD +65
1638 in the star-forming region around NGC 7129 (Matthews et al., in
preparation), IRAS 01312+6545 found in the Canadian Galactic Plane Survey,
and IRAS 06084-0611 associated with BD +30 549.  These are all examples of
"dissociating stars" with several M$_{\odot}$ of HI and a very small amount
of ionized gas (1/1000 of the HI) surrounding B stars of types ranging from
0.5 to 5.  Charles Kerton (private communication) has recently provided a
list of additional candidates (Table \ref{table:IRAS-B}).  These objects
all have HI masses of a few M$_{\odot}$.

\begin{table}
\begin{center}
\caption{\hspace*{8mm}Additional IRAS - B-star associations.
\label{table:IRAS-B}}
\vspace*{3mm}
\begin{tabular}{ccc} 
\tableline
IRAS Number & D$_{kin}$(pc) & Sp. Type \\ 
\tableline
\vspace*{-3mm}
 & & \\
01431+6232 & 6.0 & B3 \\
01448+6239 & 4.8 & B3 \\
01546+6319 & 6.0 & B3 \\
01524+6332 & 4.5 & B5 \\ 
\tableline 
\tableline
\end{tabular}
\end{center}
\end{table}

\begin{figure}
\plotone{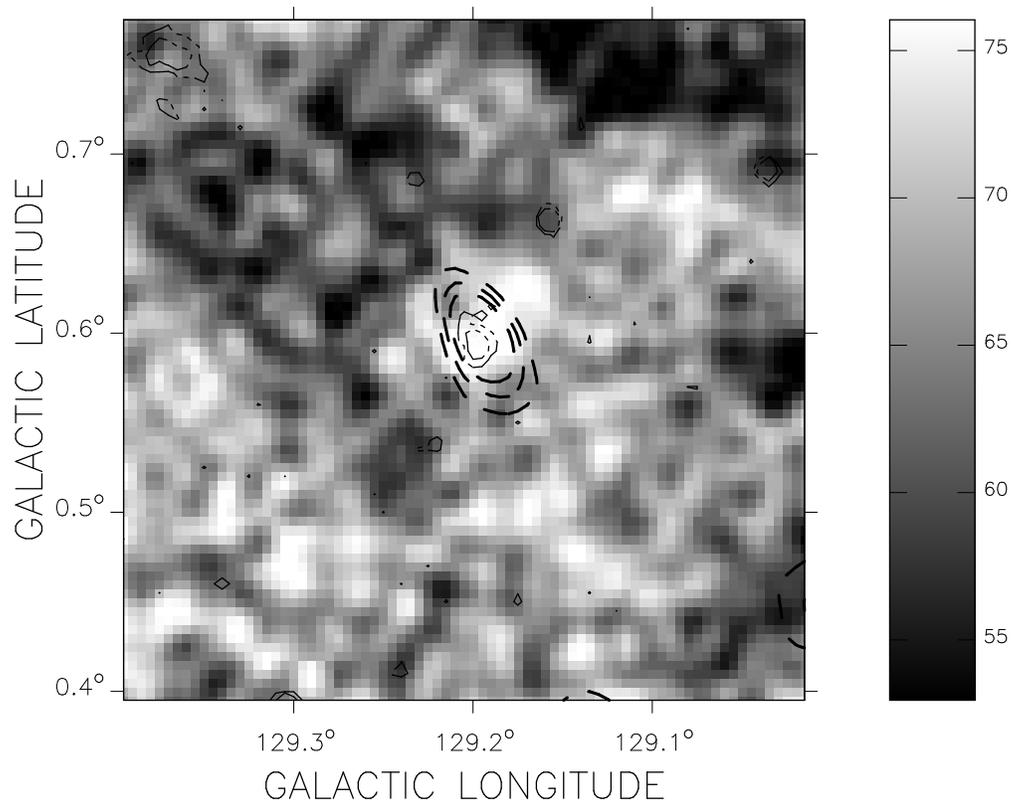}
\caption{
Emission associated with IRAS 01431+6232. The greyscale is HI
emission from 53-76 K; the outer dashed contours are CO emission at 0.4,
0.6 and 0.8 K km s$^{-1}$; and, the inner contours are MSX Band A infrared
emission at $1 \times 10^6$ and $2 \times 10^6$ Watts m$^{-2}$ sr$^{-1}$.
Figure courtesy Charles Kerton, DRAO.
\label{fig:IRAS01431+6232}
}
\end{figure}

The identification of these objects on HI maps of the Galactic plane is
facilitated by the very compact character of their HI emission.  Presently
the search criterion involves a correspondence of an HI concentration with
a compact source of IR emission, but the best way to find these objects in
the Galaxy is still under discussion.  Figure \ref{fig:IRAS01431+6232}
shows the HI and CO(1-0) emission identified with IRAS 01431+6232.

Table \ref{table:IRAS-B} is likely to be affected by observational
selection.  For instance, the HI emission from such objects which are
closer to us will be larger in angular size and more diffuse, and therefore
increasingly difficult to discern in the HI maps; on the other hand, at
distances of a few kpc, these objects are barely resolved, so more distant
HI objects will be fainter and offer less contrast with the ambient HI\@.
The concentration to mid-B spectral types may also be a selection effect;
later-type stars produce too little dissociating radiation, so not enough
HI is produced, whereas the ionizing flux from earlier-type stars destroys
the HI, creating a substantial HII region instead.

\subsection{The dissociated HI mass; a simple calculation}

The central star e.g.\ in IRAS 23545+6508 has T$_{\rm eff} \approx$ 17,000
- 18,000 K, so the rate of production of ionizing photons is ${\mathcal
R_{\rm LYC}} \approx 10^{43.3}$ photons/sec.  But the rate of production of
{\it dissociating} photons is much higher, ${\mathcal R_{\rm FUV}} \approx
10^{46.5}$ photons/sec (e.g.\ Puxley et al.\ 1990) in the range
$912 < \lambda < 1108$ \AA.

According to Dewdney et al.\ (1991) the GMC density $n_2 \approx 900$
molecules cm$^{-3}$, so the 2 $\times$ HI $\rightarrow$ H$_2$
reformation time on grains $ \tau_{\rm rec} \approx 3.3 \times 10^8/n_H$
years is $\approx 1.8 \times 10^5$ yr for $n_H = n_1 + 2n_2$, significantly
longer than the estimated present age of the B star of $ \tau_{\rm B3V}
\approx 10^4$ yr.  In this case a simple calculation can be made of the
total mass of dissociated HI around the star.  Assuming that 100\% of the
dissociating photons produced are used inside the GMC, and taking the
efficiency $\eta \approx 0.15$ for the dissociation of an H$_2$ molecule
into two HI atoms, the number $\mathcal{N}$ of HI atoms produced by the
star up to the present time is:
\begin{eqnarray*}	
\mathcal{N}({\rm HI}) & = & 2 \times \eta \times \mathcal{R}_{\rm FUV}
\times \tau_{\rm B3V}, \\
	& \approx & 3 \times 10^{57}\; {\rm HI \; atoms}, \\
{\rm M}	& \approx & 2.5\; {\rm M}_\odot {\rm \; of \; HI.}
\end{eqnarray*}
\noindent The excellent agreement with the observed value of 1.4
M$_{\odot}$ must be fortuitous, considering the approximations.  Dewdney et
al.\ (1991) used a more complete, time-dependent model of HI production
they developed themselves (Roger \& Dewdney 1992) to arrive at a value of 2
-- 3 M$_{\odot}$ of HI, again assuming a lifetime of $10^4$ yr for the
star.

Note that if the volume density of the molecular ISM in which these stars
are embedded was lower, say $n_2 \approx 100$ cm$^{-3}$ (the typical GMC),
the HI sources would be $\approx 10$ pc in size (with the same HI mass of a
few M$_{\odot}$), and would begin to merge into the generally-lumpy
distribution of Galactic HI, making them even more difficult to identify.

\section{HI associated with H$_2$ on the $\approx$ 10 pc scale}

On the $\approx 10$ pc scale in the Galaxy, confusion is a major problem,
and specific associations of HI and H$_2$ have been made only in cases
where the geometry is particularly simple, or the region is not confused
by other features along the line of sight (e.g.\ at high Galactic
latitude).

\subsection{HI and H$_2$ in interstellar cirrus clouds}

A study of the HI, CO, and IR emission from a sample of 26 isolated,
high-Galactic-latitude cirrus clouds (Reach, Koo, \& Heiles 1994) with
typical sizes 1 - 10 pc shows that the H$_2$ content of these clouds is of
the same order as the HI mass, in spite of their location above the
Galactic plane where we might expect the UV flux from the Galaxy to have
destroyed all the H$_2$.  In one well-studied case (G236+39) the inferred
mass of H$_2$ is 70 M$_{\odot}$, compared to an HI mass of 90 M$_{\odot}$.

\subsection{Warm HI ``envelopes'' around Galactic GMCs}

An analysis of observations of several Galactic GMCs (Andersson \& Wannier
1993 and references there) reveals warmed HI surface layers with
characteristic depths of $\approx 2$ pc and maximal extents of $\approx 10$
pc.  The HI production is ascribed to photodissociation.  In one
well-studied case (Andersson, Roger, \& Wannier 1992) the envelope around
the GMC called ``B5'' in the Per OB2 association has 350 M$_{\odot}$ of HI
in an envelope which is moderately dense, $\approx 35$ cm$^{-3}$, and warm,
$\approx 70$ K, and is expanding away from the cloud at approximately the
escape velocity.  Blitz \& Terndrup (quoted in Blitz 1993) have reported on
about a dozen cases of HI envelopes surrounding Galactic GMCs in or near
well-known star-forming regions (NGC 7023, S140, Per OB2, Orion, ...).  The
HI masses range from $\approx 500$ to 500,000 M$_{\odot}$.

\subsection{Another simple calculation}

How much HI could be produced by photodissociation around a typical GMC
embedded in the mean interstellar radiation field (ISRF)?  For this
estimate our initial approach assuming a very young PDR and neglecting
re-formation of H$_2$ will not be adequate.  We need a steady-state
calculation, where the HI production rate by photodissociation of H$_2$ is
balanced by the 2 $\times$ HI $\rightarrow$ H$_2$ reformation rate on grain
surfaces.  We use the model developed by Sternberg (1988) to determine the
steady-state column density of HI\@.  For standard values of ISM parameters
in the solar neighborhood we have:
\[ {\rm N}(HI) \approx 5 \times 10^{20} \times
\ln (90 \chi/n_{\rm H} + 1) \]
\noindent where:\\

\begin{tabular}{rcl}
N(HI) & = & the HI column density in atoms cm$^{-2}$, \\
$\chi$   & = & the FUV intensity relative to the local ISRF, and \\
$n_{\rm H}$      & = & the total proton volume density of the gas. \\
\end{tabular}\\

\noindent The standard GMC has a diameter of 50 pc, an average H$_2$ volume
density $\langle n_2 \rangle \approx 100$ and a total mass of $10^{5-6}$
M$_{\odot}$ (Blitz 1993).  The ISRF therefore creates a ``skin'' of HI on
the surface of this cloud with N(HI) $\approx 3.5 \times 10^{20}$ atoms
cm$^{-2}$.  Assuming the cloud is spherical, its surface area is $7 \times
10^{40}$ cm$^2$, so the warm HI envelope around the GMC will contain:\\

\begin{tabular}{rcl}
$\mathcal{N}$(HI) & $\approx$ & $3.5 \times 10^{20} \times 7 \times 10^{40}$
HI atoms, which is \\
M(HI)	& $\approx$ & $2.1 \times 10^4$ M$_{\odot}$ of HI, \\
\end{tabular}\\

\noindent ...close to the average observed HI mass in the GMC sample of
Blitz \& Terndrup. Such GMCs are apparently about 10\% HI and 90 \% H$_2$.

\section{HI associated with H$_2$ on the $\approx$ 100 pc scale}

On the $\approx 100$ pc scale in the Galaxy, confusion remains a big problem.
Unusual morphologies, correspondences in several tracers, and careful
isolation of features along the line of sight are necessary.

\subsection{HI ``tails'' and ``cones''}

\begin{figure}
\plotone{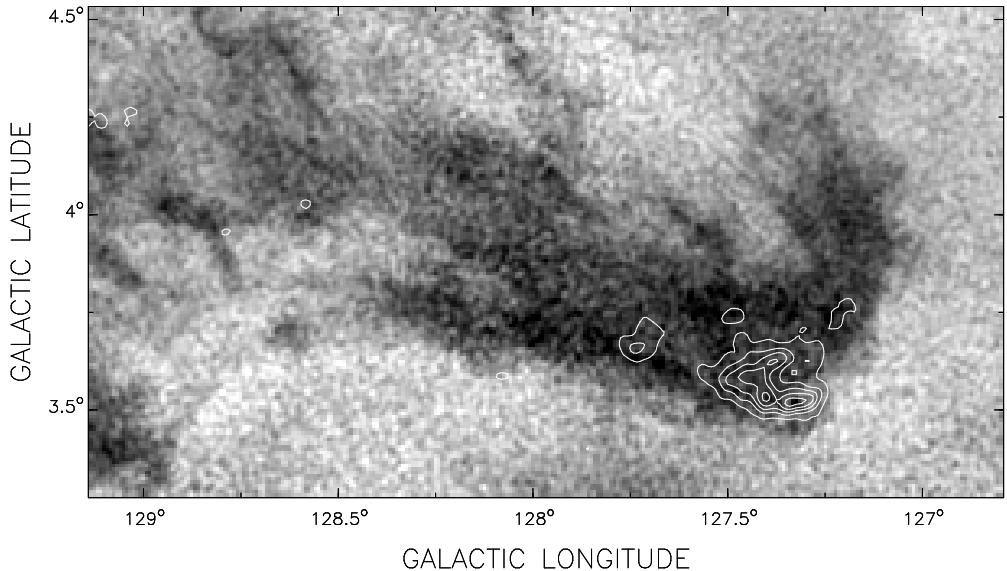}
\caption{
HI and CO(1-0) emission associated with WK-7. The greyscale is HI
emission from 0-40 K; the contours are CO(1-0) emission at 0.3, 0.8, 1.3,
1.8, 2.3, and 2.8 K km s$^{-1}$. Figure courtesy Lewis Knee, DRAO.
\label{fig:WK-7}
}
\end{figure}

The Canadian Galactic Plane Survey in HI and radio continuum, coupled with
a similar coverage in CO emission with the FCRAO radio telescope, has
revealed many features in the Galaxy which bear the signature of the
destruction of GMCs by photodissociation (e.g.  Wallace \& Knee 1999).  The
object WK-7 is a typical example (Figure \ref{fig:WK-7}), with CO emission
at the apex of an HI structure that fans out away from the CO.  For many of
these associations B stars have been found nearby.

\subsection{Maddelena's cloud}

A careful separation of features in radial velocity has permitted the
identification of a large ($\approx 50 \times 200$ pc), ``blanket'' of HI
located in the Galaxy and associated with CO emission from the GMC G216-2.5
(Williams \& Maddelena 1996) and two nearby young stars (at least one of
which is surely a B star).  This structure is especially important to the
discussion now because it is large enough to be identified in
high-resolution HI studies of nearby galaxies, and shows us the morphology
we ought to be seeking in that case.  An interesting point with this object
is that both the HI and the CO emission from this structure appear roughly
to fit into the PDR picture.  The observations give the
``excess'' HI associated with the PDR as $\approx 2 \times 10^{20}$ atoms
cm$^{-2}$ and the average CO(1-0) intensity as $\approx 6$ K km s$^{-1}$.
These two points are in rough agreement with the combined predicted CO(1-0)
(Kaufman et al.\ 1999) and HI intensity (equation given above) for an H$_2$
density of $\approx 100$ cm$^{-3}$, typical for GMCs in the Galaxy (Blitz
1993), and for an incident FUV flux of $\chi \approx 1$, which is just the
estimated ``excess'' value coming from the nearby B stars according to
Williams \& Maddelena. A more refined model is under construction
(Allen, Heaton, \& Kaufman, in preparation).

\subsection{The step to nearby galaxies}

With its $\approx 100$ pc size scale, Maddalena's Cloud provides a vital
link between the structures we have so far been discussing in the Galaxy
and the structures we can discern in the nearby galaxies with current
instrumentation.  HI ``blankets'' and ``blisters'' similar in structure to
Maddalena's Cloud and characteristic of large-scale PDR morphology have now
been found all over the disks of two nearby galaxies, M81 and M101, thanks
to the availability of FUV imaging from the ASTRO/UIT missions and
high-resolution HI images from the VLA.  The combination permits a
comparison of the FUV and HI morphologies on a $\approx 100 - 200$ pc
scale.  Although even better linear resolution would be highly desirable,
there are three important advantages of working on nearby galaxies:

\begin{itemize}

\item The line-of-sight confusion problem which plagues the Galactic
work is now virtually absent, so the features are more easily identified.

\item The distances to different structures are all very nearly the same,
so we do not have to contend with widely varying angular scales on the sky
for what are basically the same physical structures just seen at very
different distances (e.g.\ a given physical structure in the Local arm of
the Galaxy can be 10 or more times the angular size of the same structure
in the Perseus arm).  The observational selection effects are therefore
more nearly the same over the whole of the galaxy disk.

\item Observational selection effects act to enhance the utility of HI for
detecting the PDRs arising when low-density GMCs are illuminated by modest
FUV radiation fields...the HI features in this case are generally larger,
so the telescope beam filling factors are larger, and the HI is therefore
easier to detect on the 21-cm VLA maps.

\end{itemize}

\subsubsection{FUV and HI in the Sb I-II spiral M\,81}

HI ``blisters'' and ``shells'' are common in M\,81 (Allen et al.\ 1997) and
are clearly associated with nearby clusters of young stars.  These clusters
sometimes also emit H$\alpha$, but not always, whereas the association with
HI is common; B stars dominate the FUV morphology at $\lambda 150$ nm.  In
addition to the association of discrete HI and FUV features in M\,81, the
general FUV brightness correlates well with the HI brightness in spiral
arms (more on this later), indicating that HI production by a general,
scattered distribution of dissociating stars is important.

\subsubsection{FUV and HI in the Sc I spiral M\,101}

In the first really detailed, quantitative study of the HI -- FUV
association, Smith et al.\ (2000) have developed a method using the
equations given earlier in this paper to discover a new probe for the
molecular gas in galaxies.  They conclude that GMCs in M\,101 have volume
densities in the range of 30 -- 1000 cm$^{-3}$ with no clear trend from the
inner to the outer parts of the galaxy.  The large-scale and well-known
decrease in N(HI) in the inner parts of M\,101 is explained in the context of
their PDR model as a result of the increasing dust/gas ratio there.

\section{HI associated with H$_2$ on the $\approx$ 1 kpc scale}

On the $\approx 1$ kpc scale in nearby galaxies the individual PDRs can not
be resolved, and we must make do with trying to interpret the general
surface brightness distributions of FUV and HI\@.  A recent result in the
outer parts of M31 bridges the gap, providing a clear association between a
general distribution of identified B stars and the associated
(low-resolution) HI surface brightness.

\subsection{Gas, dust, and young stars in the outer disk of M\,31}

Cuillandre et al.\ (2001) have studied a field in the far outer parts
of M\,31, beyond the De Vaucouleurs radius along the major axis of the
galaxy to the south-west. Besides other results in
that paper we note the following points of interest here:

\begin{itemize}
\item There is dust mixed in with the HI on the large scale, so
we may expect that H$_2$ is also present.
\item Young stars are distributed generally with the HI\@. These
stars have the signatures (V mag, V-I colors) of B stars in the
outer parts of M31.
\item These B stars are found at radial distances of 23 to 33 kpc
(4 to 5.7 times the disk scale length in B) in areas where the galaxy
optical surface brightness is $\mu_B > 27$ mag arcsec$^{-2}$. This
is approaching the range of a low-surface-brightness galaxy.
\item Faint H$\alpha$ emission has very recently been detected from
some of the richest concentrations of bright MS stars, confirming the
interpretation that these are mostly B stars (Cuillandre,
private communication).
\item Computations of the FUV output from the census of observed B
stars and a fit of the data to PDR models is under way; preliminary
results are encouraging and give reasonable values for the densities
of the GMCs which must be the antecedents of the B stars.
\end{itemize}

\subsection{FUV and HI surface brightness correlations in M\,81}

\begin{figure}
\plotone{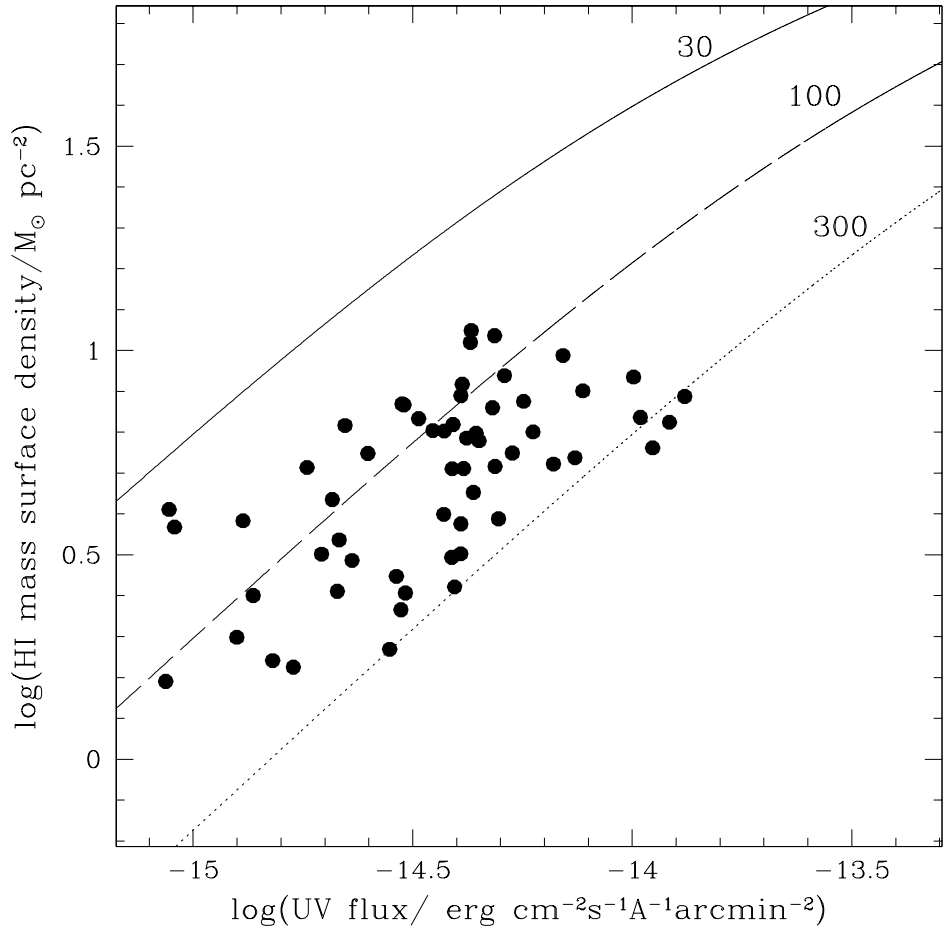}
\caption{Global comparison of HI mass surface density and FUV surface
brightness at individual locations in the galaxy M\,81 at an angular
scale of $\approx 1'$. From Emonts
et al.\ (in preparation). The curves are from a simple photodissociation
model which is only very approximate at the moment; a more complete
calculation is in progress.
\label{fig:M81-HI-FUV}
}
\end{figure}

A general correlation of VLA-HI and UIT-FUV surface brightness on the
$\approx 1$ kpc scale has very recently been shown in M81 (Emonts et al.,
in preparation).  Figure \ref{fig:M81-HI-FUV} shows this correlation when
the data is smoothed to $\approx 1'$.  The results are consistent with a
photodissociation model, although the precise parameters (e.g.\ the H$_2$
volume density) are uncertain owing to the averaging procedure on the
model.  The correlation is closely similar to that determined for a sample
of nearby galaxies from the FAUST Far-UV survey (Deharveng et al.\ 1994),
although the interpretation is quite different; this
particular result and its relation to the subject of the global star
formation rate in galaxies is discussed in more detail by Allen (2002).

\subsection{Yet another simple calculation}

What fraction of the total HI in a galaxy is maintained by the
photodissociation process we have been discussing here?  As an example,
let's take the case of NGC 4152, an Sc galaxy at D $\approx 19.5$ Mpc in
Virgo.  The FAUST Far-UV flux is $6.3 \times 10^{-14}$ ergs/cm$^2$/sec/\AA\ 
(Deharveng et al.\ 1994) at $\lambda \approx 1500$ \AA.  This corresponds
to about 1 photon/cm$^2$/sec at earth, or $\approx 4 \times 10^{52}$ Far-UV
photons/sec at the galaxy.  Let us take the fraction trapped in the galaxy
to be $f_t$.  What is the appropriate time scale?  This depends of course
on the reformation time for the process 2 $\times$ HI $\rightarrow$ H$_2$
on grains, which we mentioned briefly earlier in this talk.  If we take a
typical GMC volume density to be $n_2 \approx 5 - 50$ H$_2$ molecules
cm$^{-3}$, the reformation time is typically $3 \times 10^{6 - 7}$ yr
for standard dust parameters.  The Far-UV photon production rate in NGC
4152 then accounts for:

\[ 2 \times 0.15 \times f_t \times 4 \times 10^{52} \times 3 \times
10^{6 - 7} \times 3 \times 10^7  \approx f_t \times 10^{66 - 67}\:
{\rm HI~atoms.}\]

A typical value for $f_t \sim 0.5$, and this galaxy contains $2.08 \times
10^9$ M$_{\odot}$ of HI, or $2.5 \times 10^{66}$ HI atoms.  So from 20\% --
200\% of the HI present could be accounted for by photodissociation! This
result is startling, and perhaps even a bit outrageous; a
more precise calculation is now clearly required.

\section{Conclusions}

\begin{itemize}

\item The {\it morphology} of HI features near Far-UV sources in disk
galaxies is consistent with that expected for (low-density) PDRs.

\item The {\it quantity} of HI in these features can be calculated using
simple photodissociation physics.

\item The {\it distribution} of HI, even at faint levels, can be explained
for the most part as a steady-state photodissociation -- reformation process
of H$_2 \Leftrightarrow 2 \times $ HI, where the Far-UV photons come mainly
from non- or only weakly-ionizing B stars, and the reformation occurs on
the surfaces of dust grains.

\end{itemize}

\noindent These conclusions raise two very important questions, still
unanswered:

\begin{enumerate}

\item Just how much H$_2$ does a typical galaxy contain?

\item How much of the HI in these galaxies is primordial?

\end{enumerate}

The hints at answers provided by the work reviewed in this talk suggest
that significantly more H$_2$ is present than we currently suspect, and
that the HI clouds in the ISM of galaxies may be mostly ``processed'' gas.

\vspace{3mm}
\acknowledgments

I am grateful to my colleagues at STScI for discussions on the topics
covered in this paper, and for the invigorating research atmosphere they
have helped to create at the Institute.  Through his annual visits to
STScI, Ken has contributed very directly to that atmosphere.  It is a
special pleasure to acknowledge the intellectual stimulation and the
personal friendship which Ken has provided to me over the years.

\section*{Discussion}

\noindent {\it Silk:\,}
Your account of the origin of the HI by photodissociation should scale
to Low Surface Brightness (LSB) galaxies. What is the status of searching for 
the evidence for your ideas in these systems?

\noindent {\it Allen:\,}
The basic idea here is that B stars ought to {\it always} accompany HI, so they
ought to be present even in the LSB disks.  In general, one needs HST to
detect upper-main-sequence stars, and only a small number of LSB galaxies
are nearby enough to make this feasible.  In fact Ken and I have tried to
obtain HST time for such work, but we have not been successful.  However, I
understand that the classic LSB galaxy NGC~2915 is on the program for HST
observations in guaranteed time with the new Advanced Camera; this galaxy
is near enough, and there is enough known about the detailed distribution
of HI, that the results ought to be clear.

\noindent {\it Harding:\,}
What is the source of the primordial H$_2$ if it is the source of gas for star
formation and for HI production?

\noindent {\it Allen:\,}
This question is in the realm of cosmology and galaxy formation, and I
don't really know much about those subjects.  I hear that some of the
latest work on galaxy formation shows that H$_2$ could be formed at an early
epoch from primordial HI and H$^+$ by reactions that do not require dust.  The
reaction rates are slow, but on the other hand there is lots of time.

\noindent {\it Bosma:\,}
There is a lot of HI around the galaxy NGC 3077 in the M81 group. According
to your ideas, there should be UV emission associated with it. Did you look
for this, and did you detect anything?

\noindent {\it Allen:\,}
The UIT field containing M81 does not extend out to NGC 3077, so we can not
do the check you describe.  However, the field does indeed include some of
the HI ``streamers'' located in the far outer parts of M81 to the east, in
the general direction of NGC3077.  My student Mr.\ Emonts has recently
produced 1$\arcmin$ smoothed versions of the HI and FUV images, and there 
is indeed
detectible FUV emission associated with this ``bridge'' HI\@.  We are looking
more closely at this result, but at first sight the correlation between HI
and FUV emission seems to agree with the general trend shown in Figure 3 of
my talk.

\noindent {\it King:\,}
There's an implication here that you didn't state outright. There's got to
be a reservoir from which future stars will be made; are you saying this
is H$_2$ rather than HI?

\noindent {\it Allen:\,}
I think the conventional wisdom is that star formation, at least for stars
of masses less than a few M$_{\odot}$, requires ``processed'' gas containing 
heavy
elements and dust as the raw material.  Such ``stuff'' turns molecular
relatively quickly compared, for example, to the rotation time of a galaxy.  So
indeed it is likely that the reservoir of gas required for future
generations of star formation will be molecular, and the HI we see is a
photodissociation product of the star formation process itself.  This
suggests that galaxies have a larger reservoir of H$_2$ than we think they do,
but unfortunately I don't yet know how to calculate how much!

\end{document}